\documentclass[sigconf,screen]{acmart}

\usepackage[utf8]{inputenc}
\usepackage{multirow}
\usepackage{xcolor}
\usepackage{tcolorbox}
\usepackage{subcaption}
\usepackage{float}
\usepackage{balance}
\usepackage{flushend}
\usepackage{listings}
\usepackage{graphicx}
\usepackage{float}
\usepackage{caption}
\usepackage[normalem]{ulem}
\usepackage{microtype}
\DisableLigatures[-]{}
\colorlet{punct}{red!60!black}
\definecolor{background}{HTML}{EEEEEE}
\definecolor{delim}{RGB}{20,105,176}
\colorlet{numb}{magenta!60!black}

\lstdefinelanguage{json}{
    basicstyle=\normalfont\ttfamily\footnotesize,
    numbers=left,
    numberstyle=\scriptsize,
    stepnumber=1,
    numbersep=8pt,
    showstringspaces=false,
    breaklines=true,
    frame=lines,
    backgroundcolor=\color{background},
    literate=
     *{0}{{{\color{numb}0}}}{1}
      {1}{{{\color{numb}1}}}{1}
      {2}{{{\color{numb}2}}}{1}
      {3}{{{\color{numb}3}}}{1}
      {4}{{{\color{numb}4}}}{1}
      {5}{{{\color{numb}5}}}{1}
      {6}{{{\color{numb}6}}}{1}
      {7}{{{\color{numb}7}}}{1}
      {8}{{{\color{numb}8}}}{1}
      {9}{{{\color{numb}9}}}{1}
      {:}{{{\color{punct}{:}}}}{1}
      {,}{{{\color{punct}{,}}}}{1}
      {\{}{{{\color{delim}{\{}}}}{1}
      {\}}{{{\color{delim}{\}}}}}{1}
      {[}{{{\color{delim}{[}}}}{1}
      {]}{{{\color{delim}{]}}}}{1},
}

\AtBeginDocument{%
  \providecommand\BibTeX{{%
    \normalfont B\kern-0.5em{\scshape i\kern-0.25em b}\kern-0.8em\TeX}}}

\newcommand{\tool}{\textsc{VulCurator}}

\begin{document}

\title[\tool{}: A Vulnerability-Fixing Commit Detector]{\tool{}: A Vulnerability-Fixing Commit Detector}


\author{Truong-Giang Nguyen}
\affiliation{%
  \institution{Singapore Management University}
  \city{Singapore}
  \country{Singapore}
}
\email{gtnguyen@smu.edu.sg}

\author{Thanh Le-Cong}
\affiliation{%
  \institution{Singapore Management University}
  \city{Singapore}
  \country{Singapore}
}
\email{tlecong@smu.edu.sg}

\author{Hong Jin Kang}
\affiliation{%
  \institution{Singapore Management University}
  \city{Singapore}
  \country{Singapore}
}
\email{hjkang.2018@phdcs.smu.edu.sg}

\author{Xuan-Bach D. Le}
\affiliation{%
  \institution{University of Melbourne}
  \city{Melbourne}
  \country{Australia}
}
\email{bach.le@unimelb.edu.au }

\author{David Lo}
\affiliation{%
  \institution{Singapore Management University}
  \city{Singapore}
  \country{Singapore}
}
\email{davidlo@smu.edu.sg}


\newcommand{\giang}[1]{{\color{red}  \thetodocounter #1}}
\newcommand{\bachle}[1]{{\color{blue} Bach \thetodocounter: #1}}
\newcommand{\hongjin}[1]{{\color{blue} HJ \thetodocounter: #1}}

\setcopyright{acmlicensed}
\acmPrice{15.00}
\acmDOI{10.1145/3540250.3558936}
\acmYear{2022}
\copyrightyear{2022}
\acmSubmissionID{fse22demo-p103-p}
\acmISBN{978-1-4503-9413-0/22/11}
\acmConference[ESEC/FSE '22]{Proceedings of the 30th ACM Joint European Software Engineering Conference and Symposium on the Foundations of Software Engineering}{November 14--18, 2022}{Singapore, Singapore}
\acmBooktitle{Proceedings of the 30th ACM Joint European Software Engineering Conference and Symposium on the Foundations of Software Engineering (ESEC/FSE '22), November 14--18, 2022, Singapore, Singapore}


\begin{abstract}
Open-source software (OSS) vulnerability management process is important nowadays, as the number of discovered OSS vulnerabilities is increasing over  time. 
Monitoring vulnerability-fixing commits is a part of the standard process to prevent vulnerability exploitation. Manually detecting vulnerability-fixing commits is, however, time-consuming due to the possibly large number of commits to review.
Recently, many techniques have been proposed to \emph{automatically} detect vulnerability-fixing commits using machine learning. These solutions either: (1) did not use deep learning, or (2) use deep learning on only limited sources of information.
This paper proposes \tool{}, a tool that leverages deep learning on richer sources of information, including commit messages, code changes and issue reports for vulnerability-fixing commit classification.
Our experimental results show that \tool{} outperforms the state-of-the-art baselines up to 16.1\% in terms of F1-score. 

\tool{} tool is publicly available at \url{https://github.com/ntgiang71096/VFDetector} \ and  \url{https://zenodo.org/record/7034132\#.Yw3MN-xBzDI}, with a demo video at \url{https://youtu.be/uMlFmWSJYOE}.
\end{abstract}




\maketitle

\section{Introduction}
Open-source software (OSS) vulnerabilities can severely damage systems. An infamous example is the Equifax Data Breach\footnote{\url{https://nvd.nist.gov/vuln/detail/cve-2017-5638}}, which led to millions of cases of identity theft. 
Another example is Log4Shell\footnote{\url{https://nvd.nist.gov/vuln/detail/CVE-2021-44228}} incident, which led to many vulnerable cloud services and applications. 
For  vulnerability management, the information of vulnerabilities are collected in the  Common Vulnerabilities and Exposures (CVE)~\cite{cve} or National Vulnerability Database  (NVD)~\cite{nvd}.
OSS users can use vulnerability information such as vulnerable version(s) of a specific third-party library or how the vulnerability is fixed to make informed decisions, e.g., migrating the dependencies to invulnerable versions or patching their own client code.

Unfortunately, in practice, there is often a delay between the time a vulnerability is fixed and the time it is publicly disclosed~\cite{sabetta2018practical}, leading to a risk that OSS users are unaware of vulnerabilities in their applications. 
Therefore, OSS users would benefit from a tool that automatically detect security-relevant changes, i.e., vulnerability-fixing commits, that are not yet disclosed~\cite{sabetta2018practical, nguyen2022hermes}. 

Many existing techniques~\cite{zhou2017automated, zhoufinding, sabetta2018practical, chen2020machine, nguyen2022hermes, le2021deepcva, sawadogo2020learning, tian2012identifying} have recently  proposed solutions for automatically identifying vulnerability-fixing commits. 
Several approaches~\cite{zhou2021spi, zhoufinding, le2021deepcva, sawadogo2020learning} use deep learning, but only consider only commit messages and code changes.
Our recent work, HERMES~\cite{nguyen2022hermes}, combines information from commit messages, code changes, and issue reports, however,
uses Support Vector Machine (SVM).
In this paper, we introduce \tool{}, a tool using a deep learning  to detect vulnerability-fixing commits based on commit messages, code changes, and issue reports.
Different from previous works, \tool{} leverages BERT-based models to represent both text-based and code-based information of a commit. Specifically, we use two RoBERTa~\cite{liu2019roberta} models for commit messages and issue reports respectively, and a CodeBERT~\cite{feng2020codebert} model for code changes. 
The output probabilities from the aforementioned classifiers are aggregated using a stacking ensemble to form the final output probability. Based on the output probability, \tool{} provides a list of commits ranked by their likelihood of being vulnerability-fixing commits. 

To evaluate the performance of \tool{}, 
we conduct an empirical evaluation on two benchmarks, including the SAP dataset proposed by Sabetta et al.~\cite{sabetta2018practical} and a newly collected dataset of  TensorFlow  vulnerabilities. 
While the former contains 1,132
vulnerability-fixing and 5,995 non-vulnerability-fixing commits written in Java and Python, the latter contains 290 vulnerability-fixing and 1,535 non-vulnerability-fixing commits from TensorFlow~\cite{abadi2016tensorflow}, a well-known deep learning framework. 
We compare \tool{} with two recently proposed approaches, HERMES~\cite{nguyen2022hermes}, which uses Support Vector Machine classifiers using information from commit messages, code changes and issue reports, and VulFixMiner~\cite{zhoufinding}, a deep learning model classifying code changes from commits.
Our experiments show that \tool{} outperforms HERMES by 16.1\% and 8.5\% on the SAP and TensorFlow dataset respectively, and
\tool{} improves over VulFixMiner by 3.9\% and 4.7\%.

\section{Background and Related Work}

\textit{Vulnerability-fixing commit classification.} Vulnerability-fixing commit classification has been an active and challenging topic in software engineering research. 
Zhou et al.~\cite{zhou2017automated} use word2vec~\cite{mikolov2013efficient} to represent commit messages and forward it to a K-fold stacking model for classification. Zhou et al.~\cite{zhoufinding} fine-tuned CodeBERT to transform code changes into embedding vectors and then use one-layer neural network to classify commits. Sabetta et al.~\cite{sabetta2018practical} and Zhou et al.~\cite{zhou2021spi} proposed to train message classifier and code change classifier separately before combining them for commit classification. 
The former approach uses Support Vector Machine, while the latter uses LSTM and multi-layer CNN. 
Nguyen et al. recently proposed HERMES~\cite{nguyen2022hermes}, which uses issue reports as a third source of information using an issue classifier and an issue linker. 
The issue linker maps commits without explicitly linked issues to  best-matching issues. 




\textit{BERT-based models.} RoBERTa~\cite{liu2019roberta} is a multi-layer bidirectional Transformer model, which is trained on a large dataset of natural language. CodeBERT 
\cite{feng2020codebert}, a variant of RoBERTa, is trained on large-scale dataset consisting of bimodal data points which refer to natural language - programming language pair, and unimodal data points which refer to only programming language. 
Both RoBERTa and CodeBERT have shown to be effective in various tasks, including vulnerability-fixing classification~\cite{zhoufinding, zhou2021spi}, type inference~\cite{kazerounian2021simtyper}, program repair~\cite{mashhadi2021applying}, program analysis~\cite{le2022autopruner} or defect prediction~\cite{zhou2021assessing}.

\section{\tool{} Architecture}
\begin{figure}[htbp]
\includegraphics[width=\columnwidth]{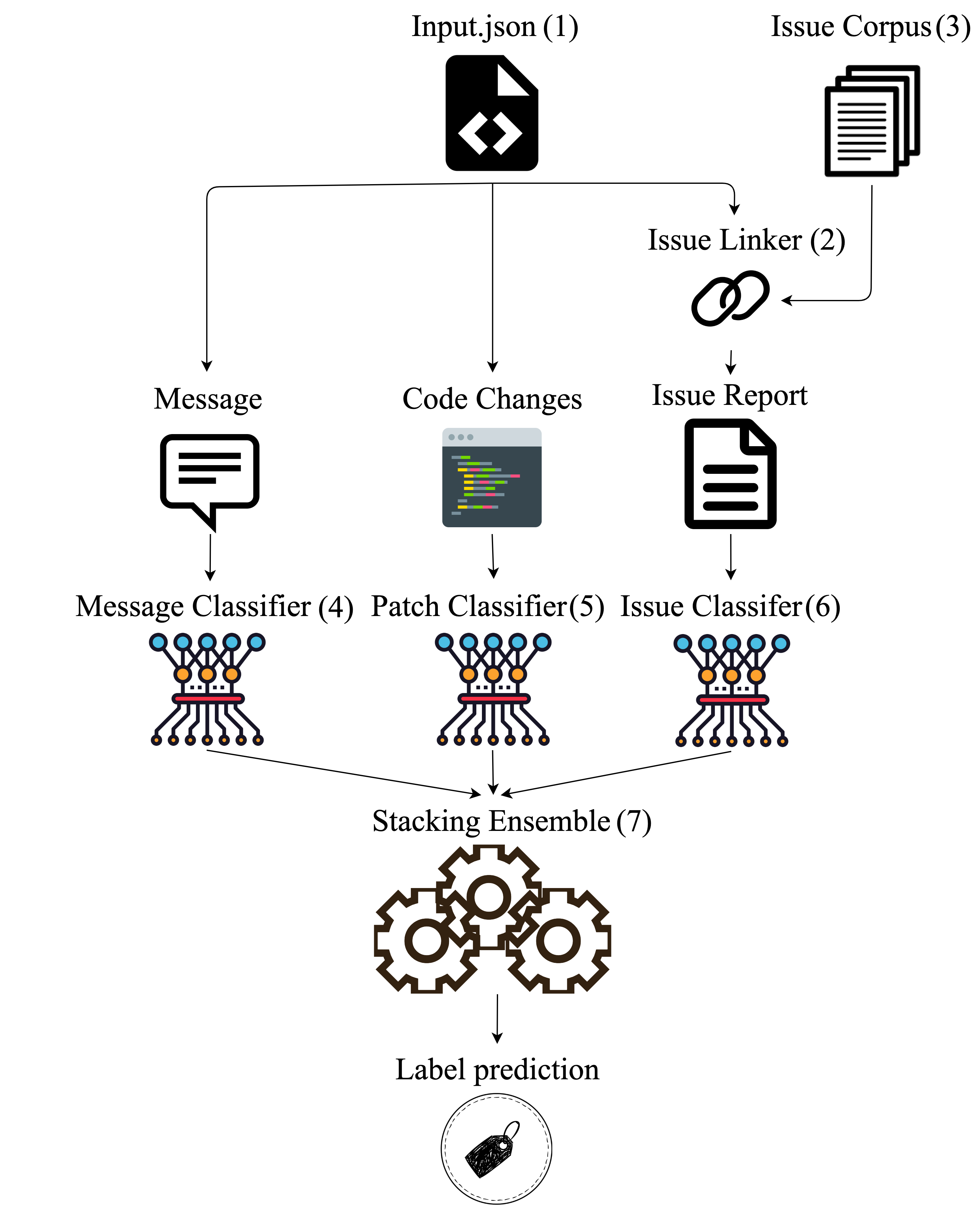}
\centering
\caption{Overview of \tool{}}
\label{fig:overview}
\end{figure}

Figure \ref{fig:overview} provides an overview of \tool{}. Our tool takes as input a JSON file \textcircled{1} containing a list of commits with their messages, code changes and linked issues. 
Note that \tool{} allows commits without explicitly linked issues. 
In these cases, \tool{} leverages an issue linker \textcircled{2}, which is built based on an issue corpus \textcircled{3} for mapping each commit to the most relevant issue in the corpus. 
Then, \tool{} feeds each type of commit information to their the corresponding classifiers, i.e. message classifier \textcircled{4}, patch classifier \textcircled{5}, or issue classifier \textcircled{6}.
Each classifier produces a probability indicating the likelihood of a commit being a vulnerability-fixing commit.  
Then, the predicted probabilities from three classifiers are combined using stacking ensemble \textcircled{7} to form the final probability. 

\vspace{1mm}
\noindent \textbf{Issue Linker.}
\tool{} first recovers commit-issue link for every commit without any corresponding issues as only a fraction of commits are explicitly linked to issue reports~\cite{sun2017frlink}.
Particularly, similar from HERMES~\cite{nguyen2022hermes}, \tool{} uses FRLink~\cite{sun2017frlink} to map each commit without any corresponding issues to its most similar issue in the input data based on a pre-defined similarity function. 
The similarity function is calculated with respect to the Term Frequency-Inverse Document Frequency (TF-IDF) of natural language terms and code terms in commit message, code changes and issue content. The TF-IDF value of every word is calculated once using TfidfVectorizer\footnote{\url{https://scikit-learn.org/stable/modules/generated/sklearn.feature_extraction.text.TfidfVectorizer.html}} and stored locally using pickle\footnote{\url{https://scikit-learn.org/stable/model_persistence.html}} for the model inference phase. 
From the findings of prior work~\cite{nguyen2022hermes}, the accuracy of commit-issue linking affects the classification performance. 
By limiting the issue linker's similarity threshold, only accurate links will be recovered. 

\vspace{1mm}
\noindent \textbf{Patch Classifier.} We use the same approach as  VulFixMiner~\cite{zhoufinding} for the patch classifier of \tool{}.
CodeBERT\footnote{\url{https://huggingface.co/microsoft/codebert-base}} is used as the core model. 
For code changes of each file, the added code and removed code version of code changes are extracted separately. 
The codes are tokenized using CodeBERT Tokenizer, and then formed as input for CodeBERT following the format below: 
\begin{equation}
    [CLS] \langle \text{rem-code} \rangle [SEP] \langle \text{added-code} \rangle [EOS]
\end{equation}
where $\mathit{rem-code}$ and $\mathit{add-code}$ are the sequence of tokens of the removed code and added code, respectively; [CLS], [SEP], [EOS] are special tokens given by CodeBERT, denoting the classification, separation and end of sequence token, respectively. The input will be forwarded to the CodeBERT to obtain an embedding vector, i.e. vector of numerical numbers, representing the semantic of code changes of each file. 
Finally, the embedding vectors are forwarded by an aggregator followed by a neural classifier to output the final probability for each commit. 

\vspace{1mm}
\noindent \textbf{Message Classifier.}
The message classifier leverages the multi-layer bidirectional Transformer model, RoBERTa~\cite{liu2019roberta}.
Specifically, a commit message is tokenized into tokens using RobertaTokenizer and then forwarded into the base version of the Roberta model\footnote{\url{https://huggingface.co/roberta-base}} and a softmax function to obtain the output probability.



\vspace{1mm}
\noindent \textbf{Issue Classifier.} Similar to the message classifier, the issue classifier also uses the base version of RoBERTa model. The model takes the commit issue's title and body as inputs, and outputs the predicted probability that the commit corresponding to the issue is for vulnerability-fixing.

\vspace{1mm}

\noindent \textbf{Stacking Ensemble and Output Prediction.} Given the output probabilities from the three aforementioned classifiers, \tool{} leverages a logistic regression model which acts as a stacking ensemble classifier to produce the final probability for each commit. 
Commits with a final probability larger than a threshold will be deemed as vulnerability-fixing commits. 
By default, the classification threshold is set as 0.5 but \tool{}  allows users to adjust the threshold (see details in Section \ref{sec:usage}).


\section{Usage} \label{sec:usage}
\subsection{Installation}

User can either clone our GitHub~\cite{github_repo} repository and install required dependencies or use our Docker image to run \tool{}~\cite{docker_img}. 
For full customization of \tool{}, 
a user can follow the following steps.

\subsection{Preparation}
\tool{} contains a built-in Issue Linker and pre-trained Classifiers, which users can directly use. However, users can build their own Issue Linker and 
Classifiers following instructions below. 


\noindent \textbf{Issue Linker.} User can customize issue corpus by providing a folder that contains files that store issue reports followed our pre-defined format (see details in our GitHub repository~\cite{github_repo}), where each issue contains issue title, issue body, and issue comments (optional). Given the corpus, users can use their own Issue Linker by using the following commands: 

\begin{center}
\texttt{python linker\_builder.py --corpus\_path <corpus\_path>}
\end{center}


\noindent \textbf{\tool{}  models.} Users can also train new classifiers for \tool{} with their own dataset by using the following command:

\begin{center}
    \texttt{python model\_builder.py --data\_path <path\_to\_data>}
\end{center}

Note that the training dataset must follow a pre-defined format, which is provided on our GitHub repository~\cite{github_repo}.

\subsection{Inference}
\tool{} provides command line interface with two modes for end-users: \textit{prediction} and \textit{ranking}.


\noindent \textbf{Input format.}
To use \tool{}, users need to prepare data following our pre-defined json format as below:

\begin{lstlisting}[language=json,firstnumber=1]
[
    {
        {
        "id": <commit_id>, 
        "message": <commit_message>,
        "issue": {
            "title": <issue_title>,
            "body": <issue_body>,
            "comments" : [<list_of_comments]
        },
        "patch": [list_of_code_change]
    },
  ...
]
\end{lstlisting}

\noindent \textbf{Prediction mode.} In prediction mode, given the input of a dataset of commits, \tool{} returns a list of likely vulnerability fixing commits along with the confidence scores. 
Although \tool{} sets the classification threshold at 0.5 by default, \tool{} allows the threshold to be adjusted with the option \texttt{--threshold}. 
Users can use the following command to obtain the results:

\begin{equation*}
    \begin{array}{lll}
    \texttt{python application.py} & \texttt{--mode} & \texttt{prediction}  \\
     & \texttt{--input} & \texttt{<input\_path>} \\
     & \texttt{--threshold} & \texttt{<threshold>} \\
     & \texttt{--output} & \texttt{<output\_path>} \\

    \end{array}
\end{equation*}

\noindent \textbf{Ranking mode.} In  ranking mode, users can input data following our format and \tool{} will output a  list of commits sorted by the probability that the commits is vulnerability-fixing. Users can use the following commands:
\begin{equation*}
    \begin{array}{lll}
    \texttt{python application.py } & \texttt{--mode} & \texttt{ranking}  \\
     & \texttt{--input} & \texttt{<input\_path>} \\
     & \texttt{--output} & \texttt{<output\_path>} \\
    \end{array}
\end{equation*}


\section{Performance Evaluation}

In this section, we investigate the following research questions:

\begin{itemize}
    \item \textit{RQ1. How effective is \tool{}?} 
    \item \textit{RQ2. How much does each classifier contribute?} 
\end{itemize}

\subsection{Experimental Setting}

\subsubsection{Dataset}
We empirically evaluate \tool{} using two datasets, the SAP dataset proposed by Sabetta et al.~\cite{sabetta2018practical} and a newly prepared TensorFlow dataset.
For each dataset, we use 80\% data for training and the remaining 20\% for testing.

\vspace{1mm}

\noindent \textbf{SAP dataset:} We evaluate our tool on the SAP dataset, which is widely used~\cite{sabetta2018practical, nguyen2022hermes}. 
The dataset contains vulnerability-fixing commits of widely used open-source projects manually-curated by SAP Security Research over a period of four years. 
Non-vulnerability-fixing commits are randomly sampled with a ratio of five non-vulnerability-fixing commits for one vulnerability-fixing commit from the same project. 
In total, the dataset contains 1,132 vulnerability-fixing and 5,995 non-vulnerability-fixing commits, in which, 37\% of the commits are explicitly linked to issues.

\vspace{1mm} 

\noindent \textbf{TensorFlow dataset:} We  introduce a new dataset with commits from TensorFlow, which is a well-known deep learning library. 
The purpose of the dataset is two-fold. First, with the increase of vulnerabilities in deep learning libraries in recent years, we would like to investigate whether \tool{} is also applicable in this domain. Second, we wish to avoid overfitting our experiments and tool design to the SAP dataset. 
To construct the dataset, we collect all vulnerability-fixing commits of TensorFlow, which are listed on National Vulnerability Database (NVD)~\cite{nvd} up until May 2022. 
We randomly sampled non-vulnerability-fixing commits from TensorFlow's repository using the same setting as Nguyen et al.~\cite{nguyen2022hermes} and Sabetta et al.~\cite{sabetta2018practical}. 
As a result, our dataset contains 290 vulnerability-fixing and 1,535 non-vulnerability-fixing commits. In this dataset, no commit is explicitly linked to an issue.




\subsubsection{Evaluation metrics} 
Similar to prior studies~\cite{tian2012identifying, zhou2021spi, nguyen2022hermes, chen2020machine}, both precision and recall are important. 
Therefore, we use F1-score, which is the harmonic mean of precision and recall, to evaluate the effectiveness of \tool{} and HERMES.

In our task, a true positive (TP) is a vulnerability-fixing commit that is correctly detected. 
A false positive (FP) is a non-vulnerability-fixing commit that is incorrectly detected as vulnerability-fixing.
A false negative (FN) is a vulnerability-fixing commit that is not detected.
Precision (P) and Recall (R) are computed as follows:
\begin{center}
  $\text {P}=\frac{\text { TP } }{\text {TP}+\text {FP}}$ 
  $\text {R}=\frac{\text { TP } }{\text {TP}+\text {FN}}$
\end{center}

Then, the F1 score is calculated as follows:
\begin{equation*}
  F1 = \frac{2(P \times R)}{P + R}\\
\end{equation*}




\subsection{Experimental Result}

\begin{table}[t]
\caption{F1 score of \tool{} and HERMES on SAP dataset. The number with the asterisk(*) denotes the result of VulFixMiner}
\label{table:rq1_sap}
\begin{center}
\begin{tabular}{|c|c|c|c|c|c|}
\hline
\textbf{Model} & \textbf{\textit{Message}} & \textbf{\textit{Issue}} & \textbf{\textit{Patch}} & \textbf{\textit{Ensemble}}  \\
\hline
HERMES & 0.67 & 0.51 & 0.60 & 0.68 \\
\hline
\textbf{\tool{}} & 0.76 & 0.65 & 0.76* & \textbf{0.79} \\
\hline
\end{tabular}
\end{center}
\end{table}


\begin{table}[t]
\caption{F1 score of \tool{} and HERMES on TensorFlow dataset. The number with the asterisk(*) denotes the result of VulFixMiner}
\label{table:rq1_tensorflow}
\begin{center}
\begin{tabular}{|c|c|c|c|c|c|}
\hline
\textbf{Model} & \textbf{\textit{Message}} & \textbf{\textit{Issue}} & \textbf{\textit{Patch}} & \textbf{\textit{Ensemble}}  \\
\hline
HERMES & 0.87 & 0.75 & 0.69 & 0.82 \\
\hline
\textbf{\tool{}} & 0.81 & 0.80 & 0.85* & \textbf{0.89} \\
\hline
\end{tabular}
\label{tab1}
\end{center}
\end{table}


\begin{figure}
    \begin{subfigure}[b]{0.45\columnwidth}
    \includegraphics[width=\textwidth]{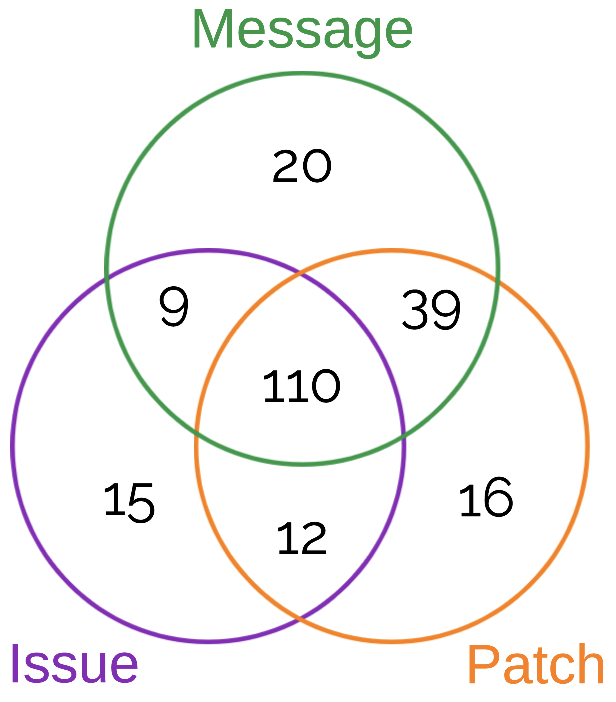}
    \centering
    \caption{SAP dataset}
    \label{fig:rq2_sap}
    \end{subfigure}
    \hfill
    \begin{subfigure}[b]{0.45
    \columnwidth}
    \includegraphics[width=\textwidth]{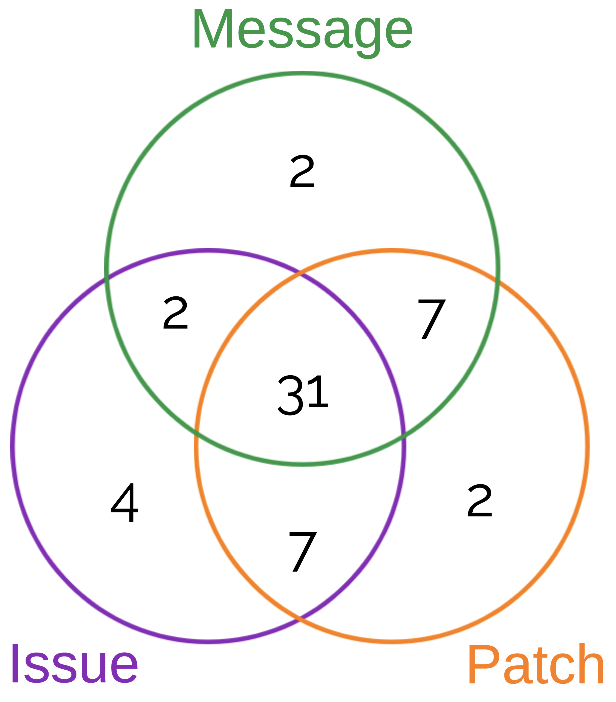}
    \centering
    \caption{TensorFlow dataset}
    \label{fig:rq2_tensorflow}
    \end{subfigure}
    \caption{Relationship between true positive cases predicted by three base classifiers of \tool{} }
    \label{fig:my_label}
\end{figure}




\subsubsection{RQ1: Effectiveness}
To answer this question, we train and test both \tool{} and HERMES on the two datasets. 
The experimental results are shown in Tables \ref{table:rq1_sap} and \ref{table:rq1_tensorflow}. On the SAP dataset, all \tool{}'s base models and the whole model outperform HERMES's. Specifically, \tool{}'s message, issue, patch classifiers and the whole model improve HERMES's counterparts by 13.4\%, 27.4\%, 26.7\%, and 16.1\% in terms of F1, respectively. 
On the TensorFlow dataset, 
while \tool{}'s message classifier has a  decrease of 6.9\% in message classifier compared to HERMES,
\tool{} issue classifier and patch classifier improves over HERMES by 6.7\% and 23.2\% respectively, leading to an overall 8.5\% improvement over HERMES. 
The experiment results suggest that \tool{} benefits from the use of pre-trained deep learning  models.

The patch classifier of \tool{} uses the same model as VulFixMiner~\cite{zhoufinding}. 
The improvement in F1 of the ensemble model over the patch classifier alone (from 0.76 to 0.79 on SAP dataset and 0.85 to 0.89 on TensorFlow dataset) shows that combining multiple sources of information allows \tool{} to outperform VulFixMiner~\cite{zhoufinding}. 
This result also validates the finding of Nguyen et al.~\cite{nguyen2022hermes} that using information from the issue tracker boosts classification performance.

\subsubsection{RQ2: Ablation Study}
\label{rq2}

We investigate if different sources of information capture different aspects of a commit. 
On the SAP dataset (Figure. \ref{fig:rq2_sap}), out of 221 discovered vulnerability-fixing commits, there are 20, 15, and 16 commits that can only be exposed by message classifier, issue classifier, patch classifier, respectively. The similar finding is also found in TensorFlow (Figure. \ref{fig:rq2_tensorflow}). The experimental results show that each classifier helps detect unique vulnerability-fixing commits.

\section{Conclusion and Future Work}

We present \tool{}, a tool for detecting vulnerability-fixing commits. 
\tool{} combines multiple sources of information such as commit messages, code changes, and issue reports in a deep learning model.
In the future, to better support security researchers in monitoring commits, 
we plan to apply explainable AI techniques~\cite{ribeiro2016should, pornprasit2021pyexplainer} to provide explanations for each prediction.

\section*{Acknowledgment}

This project is supported by the National Research Foundation, Singapore and National University of Singapore through its National Satellite of Excellence in Trustworthy Software Systems (NSOE-TSS) office under the Trustworthy Computing for Secure Smart Nation Grant (TCSSNG) award no. NSOE-TSS2020-02. Any opinions, findings and conclusions or recommendations expressed in this material are those of the author(s) and do not reflect the views of National Research Foundation, Singapore and National University of Singapore (including its National Satellite of Excellence in Trustworthy Software Systems (NSOE-TSS) office).

Xuan-Bach D. Le is supported by the Australian Government through the Australian Research Council’s Discovery Early Career Researcher Award, project number DE220101057

\balance

\bibliographystyle{ACM-Reference-Format}
\bibliography{software.bib}

\newpage

\appendix
\section{Demonstration}
This is a run-through demonstration for \tool{} using our Docker image. For manual installation, please check our GitHub repository. We also provide a demo video of \tool{} at the link \url{https://youtu.be/uMlFmWSJYOE}

\vspace{2mm}

\noindent \textbf{Step 1}: User installs \tool{} by pulling Docker image using the command:

\vspace{2mm}

\texttt{docker pull nguyentruongggiang/vfdetector:v1}

\vspace{2mm}

After a successful install, you should see a similar result to the screenshot below:

\begin{figure}[H]
    \centering
    \includegraphics[width=\columnwidth]{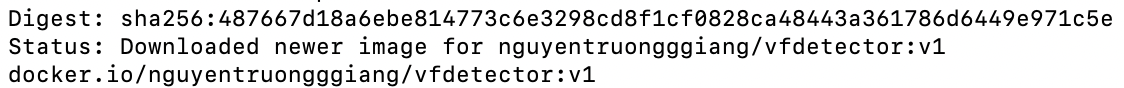}
    \caption{Installation Success}
    \label{fig:my_label}
\end{figure}

\noindent \textbf{Step 2}: Open Docker container using the command:

\vspace{2mm}

\texttt{docker run --name vfdetector -it --shm-size 16G --gpus all nguyentruongggiang/vfdetector:v1}

\vspace{2mm}

\noindent \textbf{Step 3} : Move to \tool{}'s working folder

\vspace{2mm}

\texttt{cd ../VFDetector}

\vspace{2mm}

\noindent \textbf{Step 4}: Inferring an output

\vspace{2mm}

User needs to prepare a JSON input file follow our format. Below is an example:

\begin{figure}[H]
    \centering
    \includegraphics[width=\columnwidth]{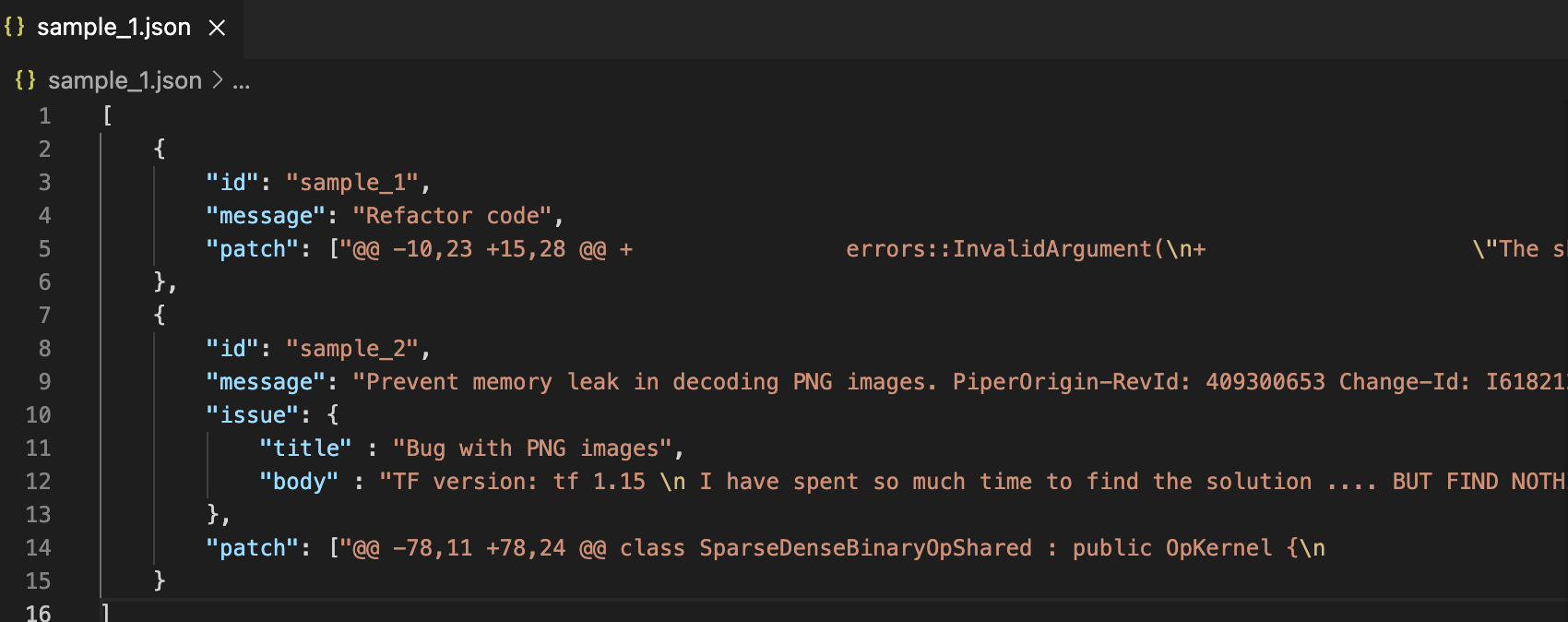}
    \caption{Input File Example}
    \label{fig:my_label}
\end{figure}

Next, run the command for either ``prediction'' mode or ``ranking'' mode:

\vspace{2mm}

\texttt{python application.py -mode prediction -input sample\_1.json -output prediction\_sample\_1.json}

\vspace{2mm}

Above is an example for "prediction" mode, which takes \textbf{sample\_1.json} as input and return \textbf{prediction\_sample\_1.json} as output.

The following output should be seen:

\begin{figure}[H]
    \centering
    \includegraphics[width=\columnwidth]{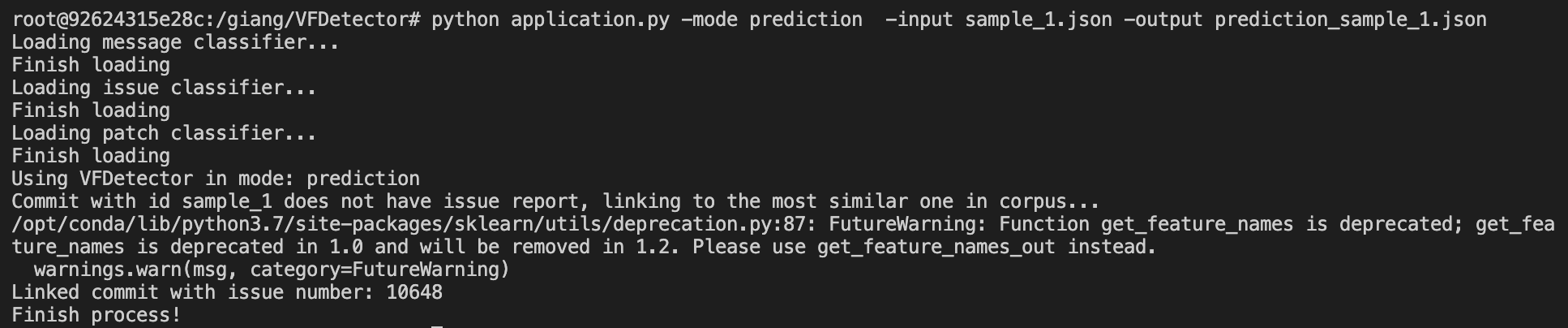}
    \caption{Screenshot for Prediction Mode}
    \label{fig:my_label}
\end{figure}

The result of the prediction is written in \textbf{prediction\_sample\_1.json}:

\begin{figure}[H]
    \centering
    \includegraphics[width=\columnwidth]{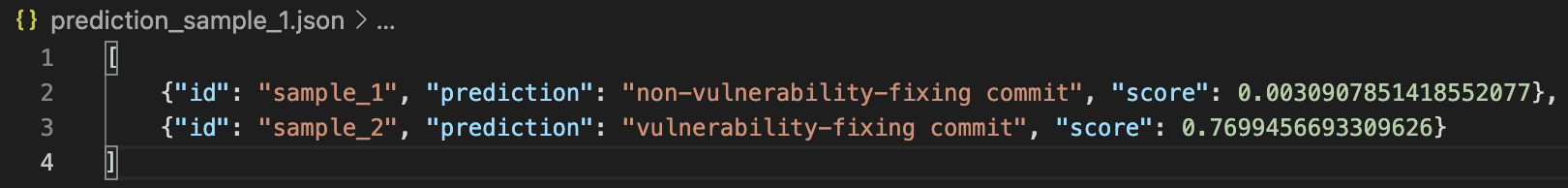}
    \caption{Example Output for Prediction Mode}
    \label{fig:my_label}
\end{figure}

Similarly, when running \tool{} in "ranking" mode, user will obtain a list of sorted commit based on the computed confidence scores similar to below:

\begin{figure}[H]
    \centering
    \includegraphics[width=\columnwidth]{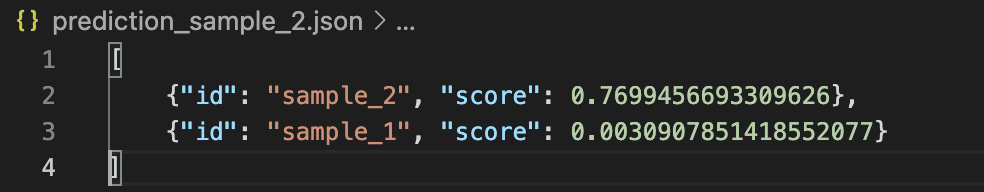}
    \caption{Example Output for Ranking Mode}
    \label{fig:my_label}
\end{figure}

\end{document}